\documentclass[aps,prd,10pt,twocolumn,superscriptaddress,amssymb,amsmath,nofootinbib,showpacs,balancelastpage]{revtex4-1}
\usepackage{graphicx,longtable,natbib,mathrsfs,color}

\newcommand{\de}{\mathrm d}

\newcommand{\om}{{\Omega_m}}
\newcommand{\ob}{{\Omega_b}}

\newcommand{\ho}{{H_0}}

\newcommand{\lcdm}{$\Lambda$CDM}

\newcommand{\fnl}{{f_\mathrm{NL}}}

\newcommand{\km}{{k_\mathrm{min}}}
\newcommand{\kM}{{k_\mathrm{max}}}

\renewcommand{\apj}{ApJ}

\newcommand{\apjl}{ApJL}
\newcommand{\aap}{A\&A}

\newcommand{\mnras}{MNRAS}
\renewcommand{\prd}{Phys. Rev. D}
\renewcommand{\prl}{Phys. Rev. Lett.}
\newcommand{\jcap}{JCAP}

\newcommand{\physrep}{Physics Reports}

\newcommand{\ijmpd}{IJMPD}
\begin{document}

\title{Neglecting Primordial non-Gaussianity Threatens Future Cosmological\\Experiment Accuracy}

\author{Stefano Camera}
\email{stefano.camera@manchester.ac.uk}
\affiliation{Jodrell Bank Centre for Astrophysics, School of Physics and Astronomy, The University of Manchester, Oxford Road, Manchester M13 9PL, UK}
\affiliation{CENTRA, Instituto Superior T\'ecnico, Universidade de Lisboa, Avenida Rovisco Pais 1, 1049-001 Lisboa, Portugal}
\author{Carmelita Carbone}
\affiliation{INAF, Osservatorio Astronomico di Brera,
Via Bianchi 46, 23807 Merate (LC), Italy}
\affiliation{INFN, Sezione di Bologna, Viale Berti Pichat 6/2, 40127 Bologna, Italy}
\author{Cosimo Fedeli}
\affiliation{INAF, Osservatorio Astronomico di Bologna, Via Ranzani 1, 40127 Bologna, Italy}
\author{Lauro Moscardini}
\affiliation{Dipartimento di Fisica e Astronomia, Alma Mater Studiorum Universit\`a di Bologna, Viale Berti Pichat 6/2, 40127 Bologna, Italy}
\affiliation{INAF, Osservatorio Astronomico di Bologna, Via Ranzani 1, 40127 Bologna, Italy}
\affiliation{INFN, Sezione di Bologna, Viale Berti Pichat 6/2, 40127 Bologna, Italy}

\date{Received \today; published -- 00, 0000}

\begin{abstract}
Future galaxy redshift surveys aim at probing the clustering of the cosmic large-scale structure with unprecedented accuracy, thus complementing cosmic microwave background experiments in the quest to deliver the most precise and accurate picture ever of our Universe. Analyses of such measurements are usually performed within the context of the so-called vanilla \lcdm\ model---the six-parameter phenomenological model which, for instance, emerges from best fits against the recent data obtained by the Planck satellite. Here, we show that such an approach is prone to subtle systematics when the Gaussianity of primordial fluctuations is concerned. In particular, we demonstrate that, if we neglect even a tiny amount of primordial non-Gaussianity---fully consistent with current limits---we shall introduce spurious biases in the reconstruction of cosmological parameters. This is a serious issue that must be properly accounted for in view of accurate (as well as precise) cosmology.
\end{abstract}


\maketitle

\section{Introduction}
The currently accepted standard model for the formation of the cosmic structure posits that the Universe underwent an early phase of accelerated expansion \citep[dubbed `inflation',][]{GU81.1} during which a random field of primordial density fluctuations was originated. Subsequent gravity-driven hierarchical growth of such density fluctuations led to the formation of galaxies, galaxy clusters and the cosmic large-scale structure (LSS). As inflation is not a fundamental theory, different classes of inflationary models predict different statistical distributions for the primordial density fluctuations \citep[see e.g.][for a comprehensive review]{BA02.2}. Naturally, structures accreting from different initial conditions will have different statistical properties. The study of such properties constitutes one of the most powerful probes for understanding the physics of the (mostly unobservable) early Universe.

One of the most general ways to quantify the statistics of primordial density fluctuations is measuring their level of non-Gaussianity. Whilst the simplest slow-roll inflationary model predicts initial conditions that are almost perfectly Gaussian, the relaxation of specific assumptions gives rise to substantial and model-dependent deviations from Gaussianity. A particularly convenient---although not unique---way to parameterise primordial non-Gaussianity (PNG) is to add a quadratic correction to the original Gaussian Bardeen's potential field \citep{SA90.1,GA94.1},
\begin{equation}
\Phi = \Phi_\mathrm{G} + \fnl\ast( \Phi^2_\mathrm{G}-\langle \Phi^2_\mathrm{G}\rangle).
\end{equation}
The quantity dubbed $\fnl$, which may be regarded as a free parameter, determines the amplitude of PNG. In the most general case, $\fnl$ may depend on both time and scale, whence the $\ast$ convolution symbol instead of ordinary multiplication; as is often done in the literature, for the sake of simplicity we here assume $\fnl$ to be scale independent.

PNG has been studied extensively over the past decade, using both data from the cosmic microwave background (CMB) and the LSS. With respect to the latter, investigations included cluster number counts, galaxy clustering, cosmic shear, LSS topology, and others \citep[see for example][and references therein]{MA00.2,GR07.1,CA08.1,GR09.1,Sartoris:2010cr,Maturi:2011am,FE11.2}. Recently, analyses of \textit{Planck} satellite data managed to severely constrain the allowed parameter space of PNG \citep{PL13.1}. Henceforth, according to a number of studies, only future LSS experiments that will be able to provide comparable constraints on $\fnl$. For example, via galaxy redshift surveys \citep{VE09.1,FE11.2,CA14.1,Raccanelli:2014awa}, in the radio continuum \citep{Maartens:2012rh,FE14.1}, with newborn techniques such as neutral hydrogen intensity mapping \citep{CA13.1} or via cross-correlation with other observables \citep{Giannantonio:2013kqa,Giannantonio:2013uqa,Raccanelli:2014kga}.

Given that many LSS cosmological tests keep finding levels of PNG that
are consistent with zero \citep[though usually with large error bars,
e.g.][]{SH13.1}, and the fact that confidence levels have been
dramatically shrunk by \textit{Planck} data, it is meaningful to ask if
PNG could be altogether ignored without significantly affecting
constraints on the other cosmological parameters. If that is the case,
the analysis of future cosmological data will be significantly
simplified. Conversely, PNG should be kept in mind
in order not to bias future cosmological constraints. This is the very
question we address in this work. Specifically, we investigate whether
we would introduce a bias in the best-fit value of other cosmological
parameters if we neglected PNG in a Universe with a small
but non-vanishing value of $\fnl$. Then, we compare this possible bias
with the statistical uncertainties predicted for future LSS
surveys. Throughout this paper we refer to a Class IV cosmological
experiment, of which the most renowned representatives are e.g.\ the Square Kilometre Array \citep[SKA,][]{SKA} at radio wavelengths, and the Dark Energy Survey \citep[DES,][]{DES}, the forthcoming European Space Agency \textit{Euclid}\footnote{\texttt{http://euclid-ec.org/}} satellite \citep[][]{LA11.2,AM12.1} and the Large Synoptic Survey Telescope \citep[LSST,][]{LSST} at optical frequencies.

As a reference model, we assume a flat \lcdm\ Universe with total
matter density (in units of the critical density) $\om=0.315$, baryon
fraction $\ob=0.0487$, dark energy equation of state CPL parameters $w_0=-1$ and $w_a=0$ \cite{CH01.1,LI02.1}, dimensionless Hubble constant
$h\equiv\ho/(100\,\mathrm{km/s/Mpc})=0.673$; the primordial power spectrum is described by its scalar spectral index $n_s=0.960$ and amplitude $A_s=2.195\times10^{-9}$ \citep{PL13.2}. We consider cosmological constraints as expected for galaxy cluster counts, clustering of galaxies and galaxy clusters, as well as their combination.

\section{Methodology}
\subsection{Modelling PNG Corrections}
The impact of deviations from Gaussianity on the abundance and clustering of the tracers of the underlying dark matter structure have been investigated by many authors obtaining either theoretical, semi-analytic or fully numerical results. Here, we summarise the most relevant aspects and refer the interested reader to e.g.~\citet{FE11.2} and references therein for additional details. PNG effects mainly concern the mass function and linear bias of dark matter haloes. These modifications involve different integrals of the gravitational potential bispectrum, $B^\Phi(\mathbf k_1,\mathbf k_2, \mathbf k_3)$. The bispectrum amplitude depends on both the amplitude $A_s$ of the gravitational potential power spectrum, $P^\Phi(\mathbf k)$, and on $\fnl$ so that
\begin{equation}
B^\Phi(\mathbf k_1,\mathbf k_2, \mathbf k_3) = f_\mathrm{NL}A_s^2\Gamma(\mathbf k_1,\mathbf k_2, \mathbf k_3).
\end{equation}
The PNG shape is determined by the dependence of $\Gamma(\mathbf k_1,\mathbf k_2,\mathbf k_3)$ upon the three momenta.

Here, we investigate the effect of some bispectrum shapes. Besides the most used local-type PNG---whose bispectrum is maximised for squeezed configurations, where one of the three momenta is much smaller than the other two---we also consider `orthogonal' PNG, so called because its configuration is nearly orthogonal to the local and equilateral shapes \citep[see][for a review]{BA04.1}. The former is known to have the heaviest impact on the clustering of the LSS, whilst the latter is nonetheless interesting because presents degeneracies with other cosmological parameters which are different to those of all other PNG types.

There are a number of prescriptions in the literature for computing PNG deviations to the abundance of dark matter haloes. Here, we follow \citet{LO08.1}, who used an Edgeworth expansion of the mass density field in order to derive a non-Gaussian generalisation of the \citet{PR74.1} mass function,  $n_\mathrm{PS}(m,z)$. We define a correction factor
\begin{equation}
\mathcal R(m,z)=\frac{n_\mathrm{PS}(m,z)}{n_\mathrm{PS}^\mathrm{(G)}(m,z)}\label{eq:rmz}
\end{equation}
by means of which one can translate any given Gaussian halo mass function $n^\mathrm{(G)}(m,z)$, computed according to one's favourite recipe, to its non-Gaussian counterpart, i.e.
\begin{equation}
n(m,z)=\mathcal R(m,z)\,n^\mathrm{(G)}(m,z).
\end{equation}
In this work, we use a Gaussian \citet{SH02.1} mass function.

Moreover, if the initial conditions for structure formation are non-Gaussian, the linear halo bias aquires  an additional scale dependence, which can be modelled as \citep{CA10.1}
\begin{equation}
b(m,z,k) = b^\mathrm{(G)}(m,z)+\beta_R(k)\sigma_m^2[b^\mathrm{(G)}(m,z)-1]^2,
\end{equation}
where  $b^\mathrm{(G)}$ is the Gaussian linear halo bias of \citet{SH01.1}. The function $\beta_R(k)$ encodes all the scale dependence of the non-Gaussian halo bias at mass $m=m(R)$, and can be written as \citep{MA08.1}
\begin{multline}
\beta_R(k) = \int_0^{+\infty}\!\!\de\xi\frac{\xi^2\mathcal M_R (\xi)}{8\pi^2\sigma_m^2\mathcal M_R(k)}\\\times\int_{-1}^{1}\!\!\de\cos\theta\frac{B^\Phi(\xi,\alpha,k)}{P^\Phi(k)}\mathcal M_R(\alpha),
\end{multline}
with $\boldsymbol\alpha=\mathbf k+\boldsymbol\xi$, $\theta$ the angle between $\mathbf k$ and $\boldsymbol\xi$, $\sigma^2_m$ the mass variance and
\begin{equation}
\mathcal{M}_R(\mathbf k)=\frac{2T(k)W_R(k)k^2}{3H_0^2\om}
\end{equation}
relating the density fluctuation field smoothed on a scale $R$ to the respective peculiar potential. $T(k)$ is the matter transfer function and $W_R(k)$ is the Fourier transform of a top-hat window function. In the case of local bispectrum shape, it has been shown that the PNG scale dependence is $\beta_R(k)\propto k^{-2}$ at large scales. For other shapes, the dependence is usually weaker \citep[see][for details]{VE09.1,FE11.2}.

In order to model the impact of PNG and other cosmological parameters on the assembly of the LSS, we make use of the well-established halo model \citep{SE00.1,CO02.2}. It is a semi-analytic framework based on the fundamental assumption that all the objects we are interested in are contained within bound dark matter haloes, so that their clustering properties can be simply expressed as a superposition of the object distribution within individual haloes and the mutual clustering properties of haloes. In this framework, galaxies are distributed within dark matter haloes according to some conditional probability distribution, $p(N_\mathrm{g}|m)$. (Note that in general this probability distribution would depend also on redshift, whereas for simplicity we ignore this dependence, unless explicitly stated.) Its first and second statistical moments, $\langle N_\mathrm{g}|m\rangle$ and $\langle N_\mathrm{g}(N_\mathrm{g}-1)|m\rangle$, respectively represent the average number of galaxies that reside within a dark matter halo of mass $m$ and the variance of that average number. A similar reasoning applies to clusters, except that it is commonly assumed that only one cluster may occupy a given dark matter halo, so that
\begin{align}
\langle N_\mathrm{c}(N_\mathrm{c}-1)|m\rangle &= 0,\\
\langle N_\mathrm{c}|m\rangle &= \Theta (m-m_\mathrm{c}),
\end{align}
where $\Theta(x)$ is the Heaviside step function and $m_\mathrm{c}$ is some---possibly redshift-dependent---mass threshold.

Hereafter, we follow \citet[][Sec.~4]{FE11.2} and by means of the halo model and of the quantities introduced above we consistently construct three-dimensional power spectra $P^{XY}(k,z)$, where $X,Y=g,c$ for galaxies and galaxy clusters, respectively. Note that, thanks to this method, we also compute the cross-correlation power spectrum between galaxies and galaxy clusters. Eventually, we calculate number counts of galaxy clusters, $N_c(z)$ as well.

Regarding the sources, we consider H$\alpha$ galaxies, which for instance will be selected by a \textit{Euclid}-like experiment. These are going preferentially to be blue star-forming galaxies, therefore we model the moments of the galaxy distribution within dark matter haloes following semi-analytic models of galaxy formation \citep{CO02.2}, which give
\begin{equation}
\langle N_\mathrm{g}|m\rangle=N_{\mathrm{g},0}\Theta(m-m_0)\left(\frac{m}{m_1}\right)^\gamma,
\end{equation}
where $N_{\mathrm{g},0} = 0.7$, $m_0 = 10^{11}h^{-1}\,M_\odot$, $m_1 =
4\times 10^{12}h^{-1}\,M_\odot$, and $\gamma =
0.8\Theta(m-m_0)$. \citet{FE11.2} showed that these choices of
parameters produce an effective galaxy bias that is in fair agreement with predictions based on semi-analytic galaxy formation models \citep{BA05.1,BO06.1} for a Class IV survey like \textit{Euclid} \citep{OR10.1}. Moreover, we set 
 \begin{equation}
\langle N_\mathrm{g}(N_\mathrm{g}-1)|m \rangle =\langle N_\mathrm{g}|m \rangle^2f(m),
\end{equation}
where the function $f(m)$ represents the deviation of the galaxy
distribution from a Poissonian, and can be modelled as
\begin{equation}
f(m) =\left\{
\begin{array}{ll}
\log^2\sqrt{\frac{m}{m_0}} & \textrm{if }m\le 10^{13}h^{-1}M_\odot\\
1 & \textrm{if }m>10^{13}h^{-1}M_\odot\\
\end{array}\right..
\end{equation}
Finally, we consider galaxy clusters that will be
photometrically selected, and for this reason we choose the minimum
cluster halo mass $m_\mathrm{c} = m_\mathrm{c}(z)$ using the \textit{Euclid} Red Book photometric selection function \citep{LA11.2}.

\section{Results}
\subsection{PNG Effects on Galaxies and Galaxy Clusters}
\begin{figure}
\centering
\includegraphics[width=\columnwidth]{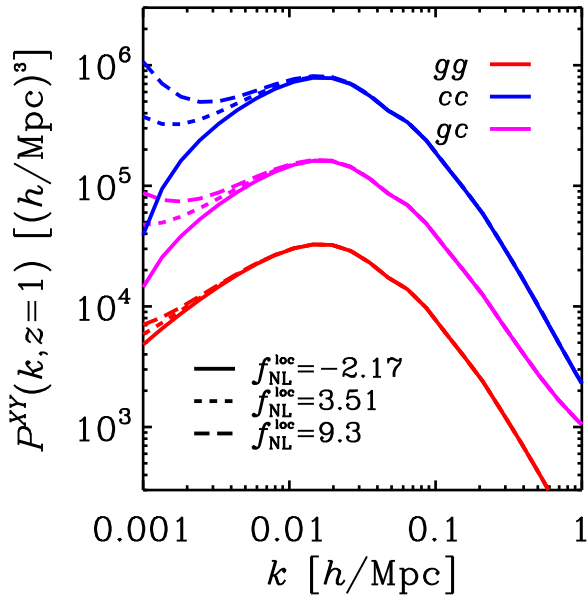}
\caption{Galaxy (red) and galaxy cluster (blue) power spectra and their cross-spectrum (magenta) at $z=1$ with $f_{\rm NL}^{\rm loc}=-2.17$, $3.51$ and $9.3$ (solid, short- and long-dashed curves, respectively).}\label{fig:powerspectra}
\end{figure}

Fig.~\ref{fig:powerspectra} illustrates the auto- and
cross-correlation power spectra of galaxies and galaxy clusters as a
function of scale at $z=1$ for three values of $\fnl$ in the local-shape scenario. Solid curves are for $f_{\rm NL}^{\rm loc}=-2.17$. We choose this particular value because, as recently remarked by \citet{CA14.1}, in \lcdm\ with slow-roll single-field inflation,
galaxy surveys should measure $\fnl\simeq-2.17$. This happens because there is a non-linear general
relativistic correction on very large scales which mimics a local PNG with $\fnl\simeq-5/3$ \citep{VE09.1,Bruni:2013qta}. This correction is derived in the CMB convention because it is based on the primordial $\Phi$. It does not affect CMB measurements of PNG, but it must be added to the local PNG parameter for LSS. The translation from CMB to LSS convention (which we adopt here) sets $f_\mathrm{NL}^\mathrm{LSS}\approx1.3f_\mathrm{NL}^\mathrm{CMB}$ \citep[see e.g.][]{FE11.2}, which eventually gives $\fnl\simeq-2.17$. The other two sets of spectra are for $f_{\rm NL}^{\rm loc}=3.51$ and $9.3$ (short- and long-dashed curves, respectively). The former has been chosen because it is the \textit{Planck} best-fit value (LSS convention) \citep{PL13.1}, whilst the latter better shows the PNG departure from the Gaussian prediction still lying within \textit{Planck} 1$\sigma$ bound.

From Fig.~\ref{fig:powerspectra} we can extract some useful information. As expected, galaxy clusters (blue curves) are more biased than galaxies (red curves), so that their power spectrum is larger. Given this, the PNG correction, which is proportional to $[b^\mathrm{(G)}-1]^2k^{-2}$, kicks in at smaller scales (larger wavenumbers) compared to galaxies, as it can be seen by looking at the different behaviour of the two curves at small $k$. Besides, we can notice that the cross-spectrum between galaxies and galaxy clusters (magenta curve set) is not merely an average of the two progenitors' spectra. Indeed, the three spectra are characterised by different scale dependences, which means that each observables carries a different piece of information about the clustering of the LSS. For instance, the different shapes at large $k$, whereby the galaxy 1-halo term carrying information on non-linear scales starts to become important, but no 1-halo term is present in the cluster power spectrum (since it is commonly assumed that only one cluster is contained inside each dark matter halo).

\begin{figure}
\centering
\includegraphics[width=\columnwidth]{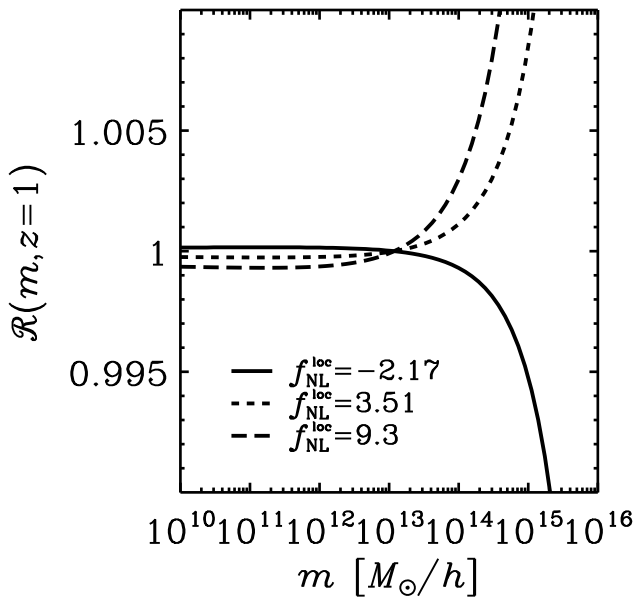}
\caption{Ratio of the non-Gaussian \citeauthor{PR74.1} mass function to the Gaussian one as a function of mass at $z=1$ with $f_{\rm NL}^{\rm loc}=-2.17$, $3.51$ and $9.3$ (solid, short- and long-dashed curves, respectively).}\label{fig:rmz}
\end{figure}

Oppositely to what happens to the power spectra, the effect of PNG on galaxy cluster counts is tinier, since it is integrated over mass and redshift. Therefore, to give a flavour of the non-Gaussian mass function, in Fig.~\ref{fig:rmz} we plot the correction factor of Eq.~\eqref{eq:rmz}, $\mathcal R(m,z)$, at $z=1$ and for the same $\fnl$ values as in Fig.~\ref{fig:powerspectra}.

\subsection{Induced Bias on Cosmological Parameters}
To estimate the bias on a set of cosmological parameters
$\{\vartheta_\alpha\}$ triggered by neglecting some amount of PNG in the data
analysis phase, we follow the Bayesian approach of
\citet{HE07.2}, based on the Fisher information matrix \citep{TE96.1}. The basic
idea is that if we try to fit against actual data a model which does
not correctly include all the relevant effects---PNG in this case---the model likelihood in parameter space will have to shift its peak in order to account for the wrong assumption. In other words, the true parameter likelihood peaks at a certain point in the full parameter space spanned by $\{\vartheta_\alpha\}\cup\{\fnl\}$; by neglecting PNG, however, we actually look at the $\fnl=0$ hypersurface, where the likelihood maximum will not in general correspond to its true value. The corresponding shift induced on the other model parameters, what we here call the bias $b(\vartheta_\alpha)$, is directly proportional to $\delta\fnl\equiv f_{\rm NL}^{\rm true}-0$ and may be computed via
\begin{equation}
b(\vartheta_\alpha)=-\left(\mathbf F^{-1}\right)_{\alpha\beta}\widehat{\mathbf F}_{\beta\fnl}\delta\fnl,\label{eq:bias}
\end{equation}
where $\mathbf F_{\alpha\beta}$ is the Fisher matrix for the wrong parameter set, $\widehat{\mathbf F}_{\alpha\beta}$ is the true Fisher matrix for the full parameter set (including $\fnl$) and $\widehat{\mathbf F}_{\beta\fnl}$ is a vector corresponding to the $\widehat{\mathbf F}$ matrix line/column relative to $\fnl$.

\begin{figure}
\centering
\includegraphics[width=\columnwidth]{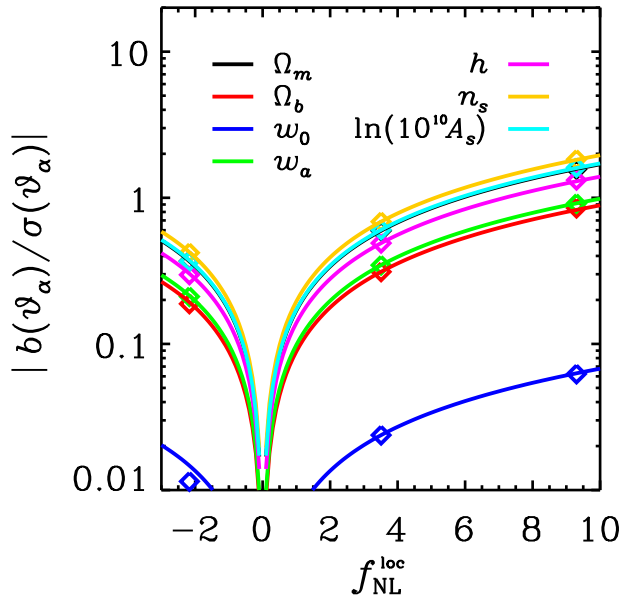}
\caption{Bias on cosmological parameters (in units of the error on the corresponding parameter) induced by neglecting PNG versus $f_{\rm NL}^{\rm loc}$ when Fisher matrices for all probes are considered.}\label{fig:bias_TOT}
\end{figure}

Details on the Fisher information matrices for galaxy and cluster
power spectra and their mutual cross-spectrum can be found in
Refs~\citep{HU08.1,FE11.2}. In the following analysis, we consider 10
redshift bins of width 0.1 centred from $z=0.9$ to $1.8$. We hold
$\kM=0.3h$ Mpc$^{-1}$ fixed to avoid the strongly non-linear r\'egime,
whilst the smallest wavenumber, $\km$, matches the largest available
scale in a given redshift bin. (To this concern, notice that general relativistic corrections on very large scales may affect the results \citep[e.g.][]{CA14.1,CA14.2}, but it has been shown that for future surveys, for instance \textit{Euclid}, their effect should be negligible
  \citep{YO13.1}.) Fig.~\ref{fig:bias_TOT} shows
$|b(\vartheta_\alpha)/\sigma(\vartheta_\alpha)|$ for the case where we
sum the Fisher matrices for all probes, i.e.\ for galaxies, clusters,
their cross-spectrum and cluster number counts. We present the bias in
units of the forecast marginal error on the corresponding parameter,
\begin{equation}
\sigma(\vartheta_\alpha)=\sqrt{(\mathbf F^{-1})_{\alpha\alpha}},
\end{equation}
better to assess the impact that such a bias will imply. The parameter set which we allow to vary is the full \lcdm\ set $\{\vartheta_\alpha\}=\{\om,\ob,w_0,w_a,h,\ln(10^{10}A_s),n_s\}$, in addition to the PNG parameter $\fnl$. Data-points refer to sampled $\fnl$ values, whereas solid curves come from the analytic expression of Eq.~\eqref{eq:bias}. Clearly,
there are parameters prone to having their `best-fit'
value shifted if we estimate them within the wrong theoretical
framework. This is the case of $\om$, $n_s$ and $A_s$,
known to be more degenerate with $\fnl$ (see correlation
coefficients in Tables~\ref{tab:loc1sigma} and \ref{tab:ort1sigma} and related discussion). On the other hand, our analysis is in good agreement with the literature, as we do not observe a significant dependence of $\sigma(\vartheta_\alpha)$ upon the assumed $\fnl$ fiducial value \citep[see e.g.][]{SE06.2,GI12.1}.

\begin{figure*}
\centering
\includegraphics[width=\columnwidth]{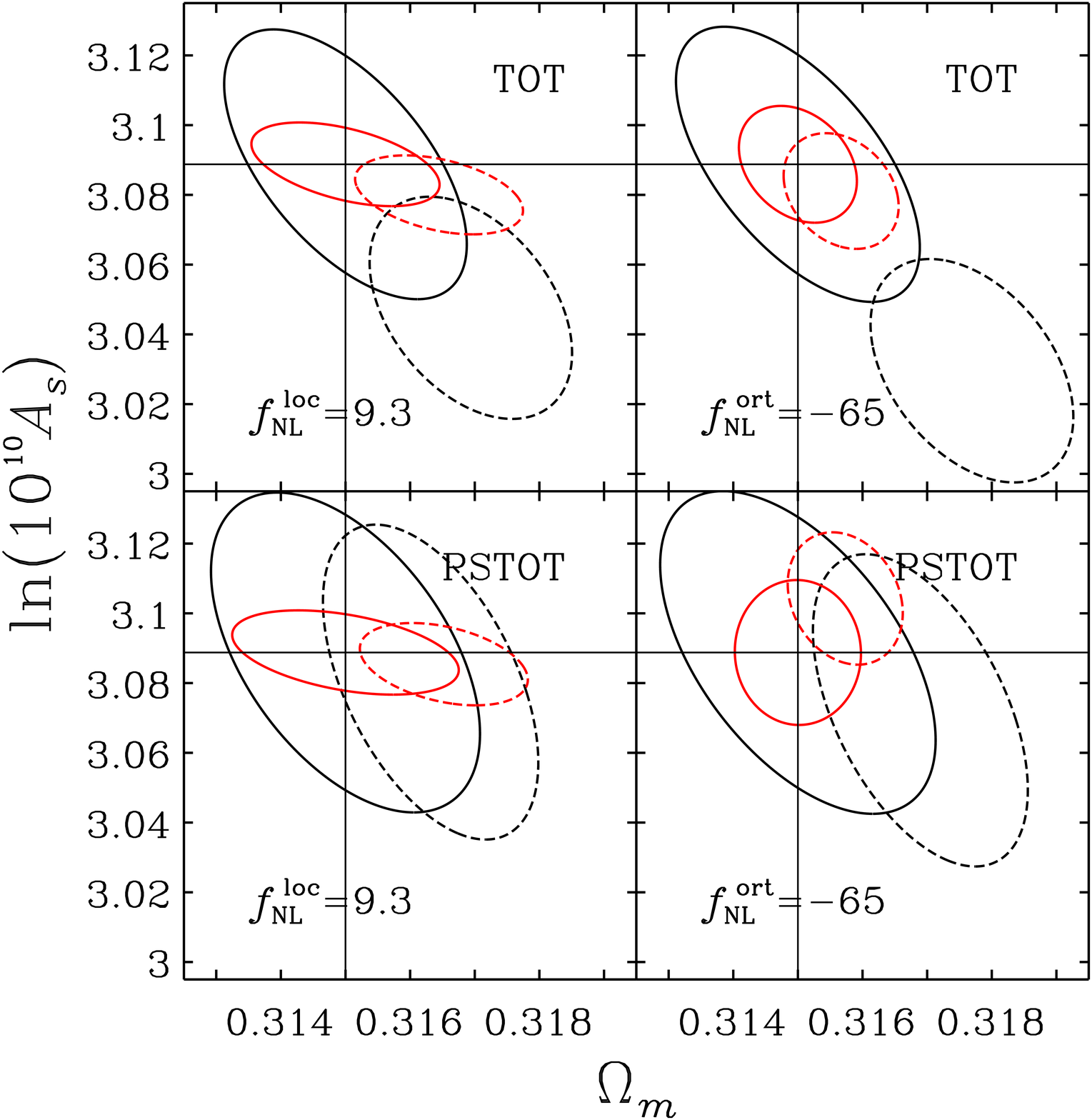}\includegraphics[width=\columnwidth]{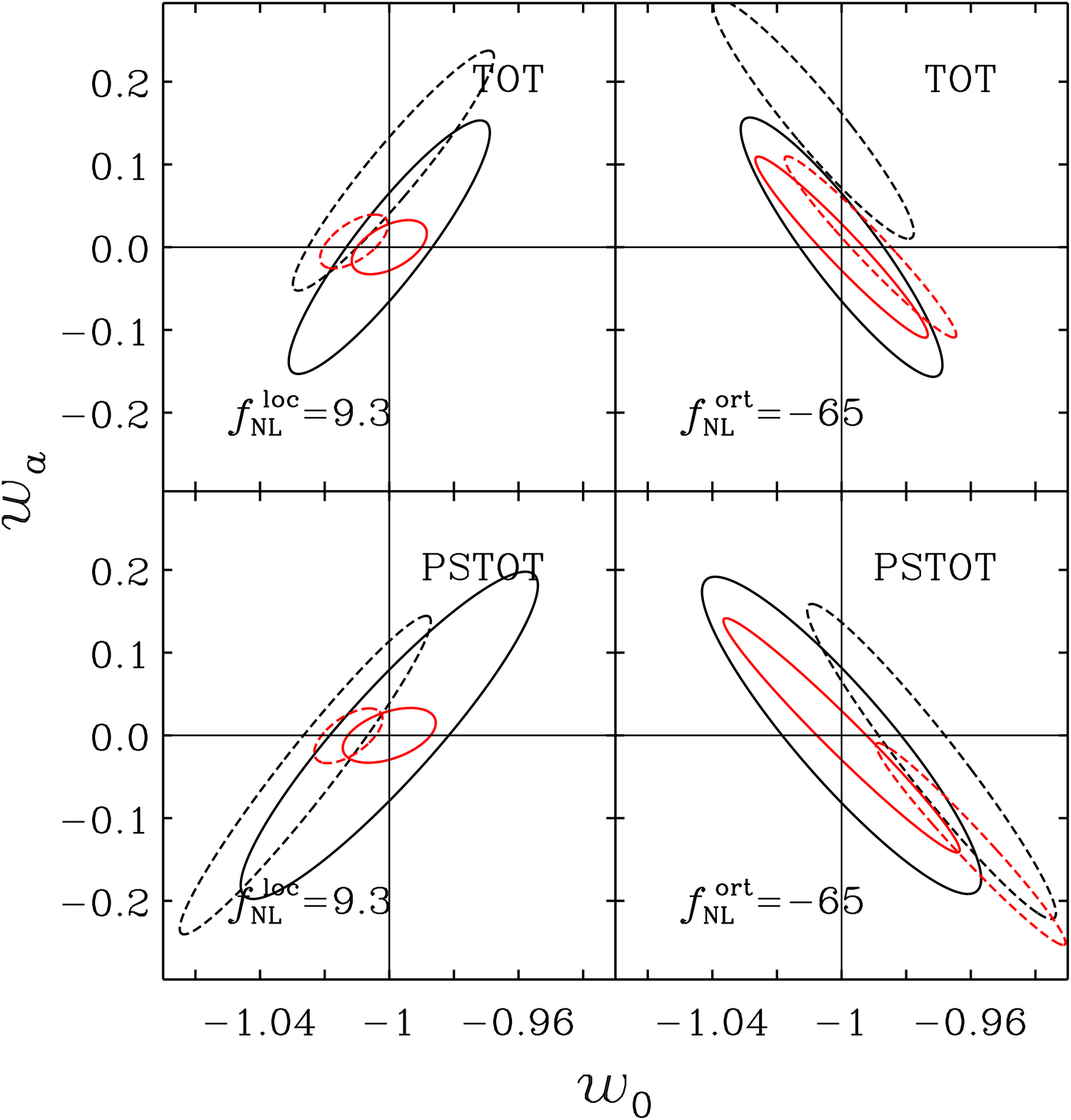}
\caption{Forecast 1$\sigma$ joint marginal contours in the  $[\om,\ln(10^{10}A_s)]$ and $(w_0,w_a)$ planes (left and right panels, respectively) for local- and orthogonal-type PNG with $f_{\rm NL}^{\rm loc}=9.3$ and $f_{\rm NL}^{\rm ort}=-65$. Solid ellipses are the true error contours, whilst dashed ellipses come from neglecting $\fnl$ in the analysis. Black and red colours respectively refer to results w/o and w/ \textit{Planck} priors on \lcdm\ parameters. Top(bottom) panels are for TOT(PSTOT).}\label{fig:ellipses}
\end{figure*}
Ultimately, this means that a blatant disregard for PNG (somehow
understandable given the stringent \textit{Planck} limits) may threaten future survey constraining power---if not by
worsening their precision, by undermining their accuracy. Surely, a
Fisher matrix approach does not fully capture the likelihood
properties on the whole parameter space. Nevertheless, we want to
emphasise that the our analysis by no means refers to some extreme situation. On the contrary, the $\fnl$ fiducial values here considered are well within
\textit{Planck} 2$\sigma$ constraints for local-type PNG.

To stress this point further on, in Fig.~\ref{fig:ellipses} we present 1$\sigma$
joint marginal contours in the $[\om,\ln(10^{10}A_s)]$ and
$(w_0,w_a)$ planes (left and right panels, respectively), with solid
lines for the true error contours from $\widehat{\mathbf F}$ and dashed contours from having neglected $\fnl$ in the Fisher analysis. This is done for local-type PNG with $f_{\rm NL}^{\rm loc}=9.3$ (left
plots in both panels) and for orthogonal-type PNG with $f_{\rm NL}^{\rm ort}=-65$ (right plots in both panels), when we consider Fisher matrices for all probes (`TOT', top plots) or only for the
combination of the three auto- and cross-spectra (`PSTOT', bottom
plots). Black and red ellipses refer to forecasts either ignoring or
including current \textit{Planck} constraints. It is clear that in
both cases, and for all the configurations and PNG choices considered
in this work, some non-negligible shift occurs. A more quantitative insight can be drawn from Tables~\ref{tab:loc1sigma} and \ref{tab:ort1sigma}, where relevant quantities on \lcdm\ cosmological parameters such as forecast marginal errors $\sigma(\vartheta_\alpha)$, $\fnl$ correlation parameters
\begin{equation}
r(\vartheta_\alpha,\fnl)=\frac{\left(\mathbf F^{-1}\right)_{\alpha\fnl}}{\sqrt{\left(\mathbf F^{-1}\right)_{\alpha\alpha}\left(\mathbf F^{-1}\right)_{\fnl\fnl}}}
\end{equation}
and normalised biases $b(\vartheta_\alpha)/\sigma(\vartheta_\alpha)$ are given.
\begin{turnpage}
\begin{table*}
\caption{\label{tab:loc1sigma}Forecast marginal errors $\sigma$, correlation parameters $r$, and (normalised) biases on \lcdm\ cosmological parameters $\vartheta$ for $f_{\rm NL}^{\rm loc}=9.3$.}
\begin{ruledtabular}
\begin{tabular}{ccccccccccccc}
\multicolumn{13}{c}{Local-type PNG ($f_{\rm NL}^{\rm loc}=9.3$)}\\
{}&\multicolumn{3}{c}{TOT}&\multicolumn{3}{c}{TOT+Planck}&\multicolumn{3}{c}{PSTOT}&\multicolumn{3}{c}{PSTOT+Planck}\\
\hline
$\vartheta$ & $\sigma$&$r$&$b/\sigma$ & $\sigma$&$r$&$b/\sigma$ & $\sigma$&$r$&$b/\sigma$ & $\sigma$&$r$&$b/\sigma$\\
\hline
$\Omega _m$ & $ 1.2\times 10^{-3}$ & $ 5.5\times 10^{-1}$ & $ 1.6$ & $ 9.6\times 10^{-4}$ & $ 4.5\times 10^{-1}$ & $ 1.5$ & $ 1.4\times 10^{-3}$ & $ 6.\times 10^{-1}$ & $ 9.6\times 10^{-1}$ & $ 1.2\times 10^{-3}$ & $ 6.7\times 10^{-1}$ & $ 1.3$ \\
$\Omega _b$ & $ 8.\times 10^{-4}$ & $ 3.\times 10^{-1}$ & $ 8.5\times 10^{-1}$ & $ 2.7\times 10^{-4}$ & $ -2.\times 10^{-1}$ & $ -6.8\times 10^{-1}$ & $ 8.6\times 10^{-4}$ & $ 2.7\times 10^{-1}$ & $ 4.4\times 10^{-1}$ & $ 2.9\times 10^{-4}$ & $ -4.1\times 10^{-1}$ & $ -8.1\times 10^{-1}$ \\
$w_0$ & $ 2.1\times 10^{-2}$ & $ 2.2\times 10^{-2}$ & $ 6.2\times 10^{-2}$ & $ 7.7\times 10^{-3}$ & $ -4.3\times 10^{-1}$ & $ -1.4$ & $ 3.\times 10^{-2}$ & $ -5.4\times 10^{-1}$ & $ -8.6\times 10^{-1}$ & $ 9.5\times 10^{-3}$ & $ -6.8\times 10^{-1}$ & $ -1.3$ \\
$w_a$ & $ 1.\times 10^{-1}$ & $ 3.2\times 10^{-1}$ & $ 9.2\times 10^{-1}$ & $ 2.2\times 10^{-2}$ & $ 9.6\times 10^{-2}$ & $ 3.2\times 10^{-1}$ & $ 1.3\times 10^{-1}$ & $ -2.3\times 10^{-1}$ & $ -3.7\times 10^{-1}$ & $ 2.2\times 10^{-2}$ & $ -1.1\times 10^{-2}$ & $ -2.2\times 10^{-2}$ \\
$h$ & $ 8.9\times 10^{-3}$ & $ 4.6\times 10^{-1}$ & $ 1.3$ & $ 1.9\times 10^{-3}$ & $ 1.6\times 10^{-1}$ & $ 5.4\times 10^{-1}$ & $ 1.\times 10^{-2}$ & $ 4.4\times 10^{-1}$ & $ 7.\times 10^{-1}$ & $ 1.9\times 10^{-3}$ & $ 1.4\times 10^{-1}$ & $ 2.8\times 10^{-1}$ \\
$\ln(10^{10}A_s$ & $ 2.6\times 10^{-2}$ & $ -5.7\times 10^{-1}$ & $ -1.6$ & $ 7.9\times 10^{-3}$ & $ -3.3\times 10^{-1}$ & $ -1.1$ & $ 3.\times 10^{-2}$ & $ -1.8\times 10^{-1}$ & $ -2.8\times 10^{-1}$ & $ 8.\times 10^{-3}$ & $ -2.1\times 10^{-1}$ & $ -4.2\times 10^{-1}$ \\
$n_s$ & $ 5.7\times 10^{-3}$ & $ -6.4\times 10^{-1}$ & $ -1.8$ & $ 2.7\times 10^{-3}$ & $ -4.7\times 10^{-1}$ & $ -1.6$ & $ 6.7\times 10^{-3}$ & $ -6.\times 10^{-1}$ & $ -9.7\times 10^{-1}$ & $ 3.1\times 10^{-3}$ & $ -6.4\times 10^{-1}$ & $ -1.2$
\end{tabular}
\end{ruledtabular}
\label{table:err_corr_bias}
\caption{\label{tab:ort1sigma}Same as Table~\ref{tab:loc1sigma} for $f_{\rm NL}^{\rm ort}=-65$.}
\begin{ruledtabular}
\begin{tabular}{ccccccccccccc}
\multicolumn{13}{c}{Orthogonal-type PNG ($f_{\rm NL}^{\rm ort}=-65$)}\\
{}&\multicolumn{3}{c}{TOT}&\multicolumn{3}{c}{TOT+Planck}&\multicolumn{3}{c}{PSTOT}&\multicolumn{3}{c}{PSTOT+Planck}\\
\hline
$\vartheta$ & $\sigma$&$r$&$b/\sigma$ & $\sigma$&$r$&$b/\sigma$ & $\sigma$&$r$&$b/\sigma$ & $\sigma$&$r$&$b/\sigma$\\
\hline
$\Omega _m$ & $1.2\times 10^{-3}$ & $-5.6\times 10^{-1}$ & $2.2$ & $6.\times 10^{-4}$ & $-2.2\times 10^{-1}$ & $1.1$ & $1.4\times 10^{-3}$ & $-6.3\times10^{-1}$ & $1.3$ & $6.4\times 10^{-4}$ & $-4.1\times 10^{-1}$ & $1.1$ \\
$\Omega _b$ & $8.\times 10^{-4}$ & $-2.8\times 10^{-1}$ & $1.1$ & $2.3\times 10^{-4}$ & $-4.6\times 10^{-2}$ & $2.3\times 10^{-1}$ & $8.6\times 10^{-4}$ & $-2.8\times 10^{-1}$ & $6.\times 10^{-1}$ & $2.3\times 10^{-4}$ & $3.2\times 10^{-2}$ & $-8.9\times 10^{-2}$ \\
$w_0$ & $2.1\times 10^{-2}$ & $1.1\times 10^{-1}$ & $-4.2\times 10^{-1}$ & $1.8\times 10^{-2}$ & $-1.\times 10^{-1}$ & $5.1\times 10^{-1}$ & $2.8\times10^{-2}$ & $-4.5\times 10^{-1}$ & $9.8\times 10^{-1}$ & $2.4\times 10^{-2}$ & $-6.\times 10^{-1}$ & $1.7$ \\
$w_a$ & $1.\times 10^{-1}$ & $-3.8\times 10^{-1}$ & $1.5$ & $7.2\times 10^{-2}$ & $-9.2\times 10^{-4}$ & $4.6\times 10^{-3}$ & $1.3\times 10^{-1}$ & $1.2\times10^{-1}$ & $-2.5\times 10^{-1}$ & $9.3\times 10^{-2}$ & $5.1\times 10^{-1}$ & $-1.4$ \\
$h$ & $8.9\times 10^{-3}$ & $-4.5\times 10^{-1}$ & $1.8$ & $1.8\times 10^{-3}$ & $-2.8\times 10^{-1}$ & $1.4$ & $1.\times 10^{-2}$ & $-4.5\times 10^{-1}$ & $9.7\times 10^{-1}$ & $1.9\times 10^{-3}$ & $-3.6\times 10^{-1}$ & $9.9\times 10^{-1}$ \\
$\ln(10^{10}A_s)$ & $2.6\times 10^{-2}$ & $5.8\times 10^{-1}$ & $-2.3$ & $1.1\times 10^{-2}$ & $1.4\times 10^{-1}$ & $-6.9\times 10^{-1}$ & $3.1\times 10^{-2}$ & $2.5\times 10^{-1}$ & $-5.4\times 10^{-1}$ & $1.4\times 10^{-2}$ & $-4.1\times 10^{-1}$ & $1.1$ \\
$n_s$ & $5.7\times 10^{-3}$ & $6.3\times 10^{-1}$ & $-2.4$ & $2.\times 10^{-3}$ & $2.6\times 10^{-1}$ & $-1.3$ & $6.7\times 10^{-3}$ & $6.1\times 10^{-1}$ & $-1.3$ & $2.1\times 10^{-3}$ & $3.6\times 10^{-1}$ & $-9.9\times 10^{-1}$ \\
\end{tabular}
\end{ruledtabular}
\label{table:err_corr_bias}
\end{table*}
\end{turnpage}

A major point emerging from this analysis is that, even though orthogonal-type PNG deviations
from the Gaussian prediction have a much smaller impact upon the
clustering of the LSS compared to PNG with local shape, current
constraints on orthogonal PNG are consequently looser. In particular,
\textit{Planck} data \citep{PL13.1} agrees with $f_{\rm NL}^{\rm ort}=-33\pm51$ (LSS
convention). That is to say, the value  $f_{\rm NL}^{\rm
  ort}=-65$ we here assume is well within \textit{Planck} 1$\sigma$ bounds. Nonetheless, if it were the true value and we neglected it, we would miss the true likelihood peak by more than 1$\sigma$---which is intolerable for the aims of future cosmological experiments.

\section{Conclusions}
In this paper, we investigated the impact of neglecting PNG when performing parameter reconstruction for an idealised representative of next generation Class IV cosmological experiments. Specifically, we considered a spectroscopic galaxy redshift survey along the lines of the European Space Agency \textit{Euclid} satellite. This allowed us to compute galaxy and galaxy cluster three-dimensional power spectra, as well as their cross-spectrum and galaxy cluster number counts, in a fully consistent way within the halo model framework.

Hence, we estimated the bias on the reconstruction of standard \lcdm\ cosmological parameter induced by disregarding PNG in the analysis. This has been done in a Bayesian Fisher matrix perspective, by considering the \lcdm\ vanilla cosmological model as a subspace (in parameter space) of a \lcdm\ Universe with PNG. In other words, we recover the concordance cosmological model if we restrict the parameter space to the $\fnl=0$ hypersurface. By doing so, the peak of the parameter likelihood
on the hypersurface does not, in general, correspond to its true peak in the full parameter space---if $\fnl$ is nonzero and it is not completely uncorrelated to the other cosmological parameters.

Our major results are summarised in Tables~\ref{tab:loc1sigma} and \ref{tab:ort1sigma} and in Fig.~\ref{fig:ellipses}. In particular, we found that an incorrect treatment of PNG in the data analysis will undermine the experimental accuracy on the reconstruction of some cosmological parameters. For example, the best-fit value of the dark energy parameters $w_0$ will be biased by more than one standard deviation, if local-type PNG is in fact present with a value of $\fnl$ consistent with 1$\sigma$ \textit{Planck} constraints. This is mainly due to the high precision of oncoming surveys, which will provide us with very tight constraints on the \lcdm\ model parameters.  Indeed, if on the one hand their expected allowed regions in parameter space will only slightly shrink by neglecting $\fnl$ in the analysis (as known in the literature), on the other hand the small but non-negligible degeneracy with $\fnl$ will cause a shift of their reconstructed best-fit value. To avoid this, it appears clear that PNG has to be consistently accounted for.

Lastly, we emphasise that, albeit we adopt the specifics of a \textit{Euclid}-like survey as a reference experiment, our findings should be regarded as potential systematics for the whole class of future, high-precision galaxy surveys, such as DES, LSST and the SKA.

\acknowledgements
We thank Alan Heavens, Roy Maartens and M\'ario G. Santos for clarifications and comments, as well as and Ben Granett for further support. SC acknowledges support from FCT-Portugal under Post-Doctoral Grant No.\ SFRH/BPD/80274/2011 and from the European Research Council under the EC FP7 Grant No.\ 280127. CC acknowledges financial support from the INAF Fellowships Programme 2010 and from the European Research Council through the Darklight Advanced Research Grant (No.\ 291521). CF has received funding from the European Commission Seventh Framework Programme (FP7/2007-2013) under grant agreement No.\ 267251. LM acknowledges financial contributions from contracts ASI/INAF No.\ I/023/12/0 `Attivit\`a relative alla fase B2/C per la missione Euclid', PRIN MIUR 2010-2011 `The dark Universe and the cosmic evolution of baryons: from current surveys to Euclid', and PRIN INAF 2012 `The Universe in the box: multiscale simulations of cosmic structure'. SC also wishes to thank Libellulart Officina DiSegni for hospitality during the development of this work.

\bibliographystyle{apsrev4-1}
\bibliography{../master}
\end{document}